\documentclass[sigconf]{acmart}
\acmVolume{1}
\acmNumber{1}
\acmArticle{}
\acmISBN{}
\acmDOI{}
\copyrightyear{2026}
\setcopyright{rightsretained}
\begin{document}

\title{DDSP-QbE++: Improving Speech Quality for DDSP-based Speech Anonymisation for Atypical Speech}

\author{Suhita Ghosh}

\email{suhita.ghosh@ovgu.de}
\orcid{0000-0002-5553-585X}

\affiliation{%
  \institution{Artificial Intelligence Lab, \\
    Otto-von-Guericke University}
  \city{Magdeburg}
  \country{Germany}
}

\author{Yamini Sinha}

\email{yamini.sinha@ovgu.de}
\orcid{0000-0002-6312-9827}

\affiliation{%
  \institution{MDS-IIKT\\
    Otto-von-Guericke University}
  \city{Magdeburg}
  \country{Germany}
}

\author{Sebastian Stober}
\email{stober@ovgu.de}
\affiliation{%
  \institution{Artificial Intelligence Lab, \\
    Otto-von-Guericke University}
  \city{Magdeburg}
  \country{Germany}
}


\begin{abstract}
  Differentiable Digital Signal Processing (DDSP) pipelines for voice conversion rely on subtractive synthesis, where a periodic excitation signal is shaped by a learned spectral envelope to reconstruct the target voice. In DDSP-QbE, the excitation is generated via phase accumulation, producing a sawtooth-like waveform whose abrupt discontinuities introduce aliasing artefacts that manifest perceptually as buzziness and spectral distortion, particularly at higher fundamental frequencies. We propose two targeted improvements to the excitation stage of the DDSP-QbE subtractive synthesizer. First, we incorporate explicit voicing detection to gate the harmonic excitation, suppressing the periodic component in unvoiced regions and replacing it with filtered noise, thereby avoiding aliased harmonic content where it is most perceptually disruptive. Second, we apply Polynomial Band-Limited Step (PolyBLEP) correction to the phase-accumulated oscillator, substituting the hard waveform discontinuity at each phase wrap with a smooth polynomial residual that cancels alias-generating components without oversampling or spectral truncation. Together, these modifications yield a cleaner harmonic roll-off, reduced high-frequency artefacts, and improved perceptual naturalness, as measured by MOS. The proposed approach is lightweight, differentiable, and integrates seamlessly into the existing DDSP-QbE training pipeline with no additional learnable parameters.
\end{abstract}


\begin{teaserfigure}
  \includegraphics[width=\textwidth]{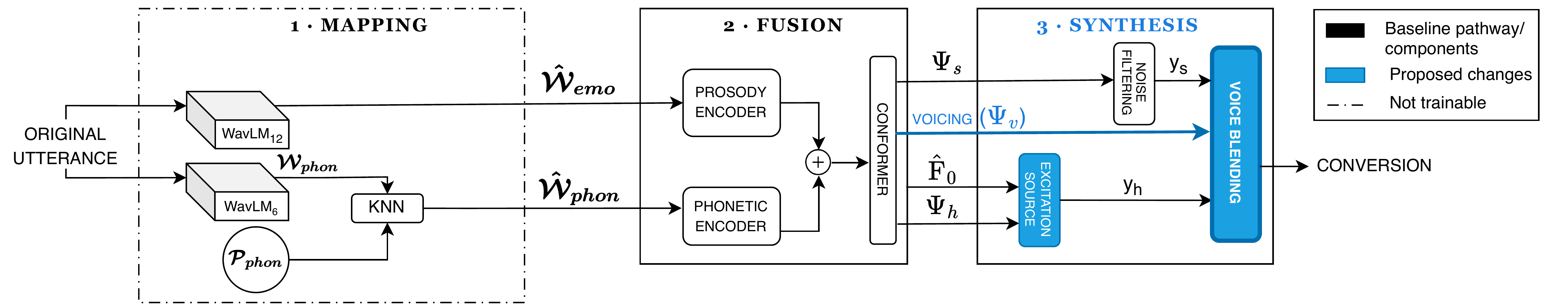}
  \caption{Overview of the DDSP-QbE++ pipeline. The system comprises three stages: 
(1)~\textbf{Mapping} -- source features are extracted via WavLM and matched 
to a target phonetic pool using KNN; (2)~\textbf{Fusion} -- emotion and 
phonetic embeddings are fused via a Conformer; (3)~\textbf{Synthesis} -- 
the fused representation drives a DDSP subtractive synthesiser with 
PolyBLEP-corrected excitation source and voicing-gated harmonic/noise 
blending. Here, voiced ($y_s$) and unvoiced ($y_h$) streams are combined 
via \textbf{Voice Blending} to produce the final conversion. Blue components 
indicate proposed modifications to the baseline DDSP-QbE pipeline.}
  \label{fig:architecture}
\end{teaserfigure}

\received{16 March 2026}

\maketitle

\section{Introduction}
Voice conversion (VC) converts a source utterance to sound like a different \textit{target} speaker while preserving its linguistic content.
In the context of pathological speech, this objective extends beyond speaker identity transfer: preserving the paralinguistic characteristics of the source, including dysfluency patterns, prosodic irregularities, and disorder-specific spectral cues, which is equally critical~\cite{yamagishi2024survey,hua2024emotional}.
Losing pitch or spectral cues can obscure clinically relevant information, degrade downstream tasks such as automatic speech recognition for assistive technologies, and reduce user experience in applications designed for speakers with motor speech disorders or voice pathologies~\cite{rajesh2025enhancement}.

Recent VC research increasingly favours non-parallel methods, avoiding the need for costly parallel source-target recordings~\cite{sisman2020overview}, a particularly important constraint for pathological speech, where matched corpora are scarce, and speaker variability is high.
One line of work uses variational autoencoders~\cite{tang2022avqvc,ghosh2024improving}, which disentangle speaker and content representations through reconstruction loss for linguistic preservation and regularisation loss (e.g., KL-divergence) for latent space structuring.
Another line explores generative adversarial networks~\cite{cycleGAN,walczyna2023overview}, adopting cycle-consistency losses for non-parallel training.
These methods take the source acoustic features (e.g., mel-spectrogram), transform them into a target-conditioned version using target-speaker embeddings, and use a vocoder to synthesise the final waveform.
However, target speaker embeddings often capture emotion and leak prosodic cues, overriding the source paralinguistic characteristics~\cite{das2023starganvcemotionpreservingvoice}, a critical failure mode when the source speaker's dysfluency or affective state must be retained.

More recently, matching-based methods such as KNN-VC~\cite{baas2023voiceconversionjustnearest} bypass explicit learning of speaker or phonetic embeddings by leveraging self-supervised representations (e.g., WavLM~\cite{chen2022wavlm}) and mapping source phonetic features to the target via K-nearest neighbour (KNN) search.
However, self-supervised features are not trained to disentangle linguistic, speaker, and pathology-related factors, so matched target features often carry unintended paralinguistic cues, causing \textit{disorder leakage}, a particular concern when the conversion target is a healthy or anonymised voice.
DDSP-QbE~\cite{ddspqbe} mitigates this by incorporating explicit emotion embeddings from the source, making it a strong candidate for pathological speech VC where preserving expressive and dysfluency-related prosody is essential.
However, the subtractive synthesiser in DDSP-QbE generates its periodic excitation via naive phase accumulation, introducing aliasing artefacts that manifest as buzziness and spectral distortion, perceptual degradations that are especially problematic for pathological voices, whose spectral fine structure carries disorder-diagnostic information.

\section{Method}
\subsection{Baseline: DDSP-QbE}
Our framework builds on DDSP-QbE~\cite{ddspqbe}, a state-of-the-art matching-based VC system.  
Given a source utterance $\mathcal{X}$, phonetic embeddings $\mathcal{W}_{\text{phon}}$ and emotion embeddings $\mathcal{W}_{\text{emo}}$ are extracted from the 6\textsuperscript{th} and 12\textsuperscript{th} WavLM layers, respectively.  
For each target speaker $s_t$, a reference pool $\mathcal{P}^{s_t}_{\text{phon}}$ is formed from all phonetic embeddings of $s_t$.  

During training, \textit{same-speaker reconstruction} is used: each source phonetic embedding in $\mathcal{W}_{\text{phon}}$ is matched to its $K$ nearest neighbours in the same-speaker pool $\mathcal{P}^{s_s}_{\text{phon}}$ to obtain $\hat{\mathcal{W}}_{\text{phon}}$. The resulting $\hat{\mathcal{W}}_{\text{phon}}$ is then fused with $\mathcal{W}_{\text{emo}}$ and passed to a differentiable subtractive synthesiser~\cite{ddspqbe} for waveform generation.
 
At inference, cross-speaker conversion is carried out: each source embedding in $\mathcal{W}_{\text{phon}}$ is matched to its $K$ nearest neighbours in the target pool $\mathcal{P}^{s_t}_{\text{phon}}$, and the averaged embeddings form $\hat{\mathcal{W}}_{\text{phon}}$.  
The source emotion embeddings $\mathcal{W}_{\text{emo}}$ are combined with $\hat{\mathcal{W}}_{\text{phon}}$ and synthesised into the final waveform.

\subsection{Proposed Changes}
\subsubsection{PolyBLEP-Based Anti-Aliasing for the Harmonic Oscillator}
The harmonic synthesizer in DDSP-QbE generates periodic waveforms by superimposing sinusoidal partials scaled by predicted harmonic amplitudes. In practice, the synthesis backend approximates this via a sawtooth-shaped excitation signal whose harmonics are modulated by a learned spectral envelope. A naive sawtooth constructed by direct phase accumulation introduces aliasing whenever harmonic content folds back above the Nyquist frequency - a particularly acute problem in high-pitched speakers or when fundamental frequency ($F_0$) varies rapidly, as is common in pathological or expressive speech.
We address this with a Polynomial Band-Limited Step (PolyBLEP)~\cite{valimaki2005discrete} correction applied directly to the phase-accumulated oscillator output before downstream spectral shaping. PolyBLEP replaces the instantaneous discontinuity of each waveform edge with a polynomial residual that cancels the alias-generating components, without requiring costly oversampling or hard spectral truncation.

\paragraph{PolyBLEP Residual Formulation}
Let $\phi[n] \in [0, 1)$ denote the normalised phase of the oscillator at discrete time $n$, and let $\Delta\phi[n] = F_0[n] / F_s$ be the instantaneous phase increment.
The raw sawtooth output is:
$x_{\text{saw}}[n] = 2\phi[n] - 1$.
A discontinuity occurs at each phase wrap $\phi[n] \to 0$.
We define the PolyBLEP residual $r(\phi, \Delta\phi)$ as a piecewise polynomial correction applied in the vicinity of each wrap:

\begin{equation}
r(t) = \begin{cases}
t^2 + 2t + 1,  -1 \leq t < 0 \\
-t^2 + 2t - 1,  0 \leq t < 1 \\
0, & \text{otherwise}
\end{cases}
\end{equation}

where $t{=}(\phi \bmod 1) / \Delta\phi$ is a normalized time offset from the discontinuity. The corrected sawtooth is then:

\begin{equation}
    x_{\text{blep}}[n] = x_{\text{saw}}[n] - r\!\left(\frac{\phi[n]}{\Delta\phi[n]}\right) + r\!\left(\frac{\phi[n] - 1}{\Delta\phi[n]}\right)
\end{equation}

This correction is differentiable almost everywhere and has negligible computational overhead relative to the sinusoidal additive synthesis baseline.

\paragraph{Integration into the DDSP-QbE Pipeline}
In DDSP-QbE, the harmonic synthesizer receives frame-level predictions of $F_0$, harmonic amplitudes $\{a_k\}$, and a noise magnitude $\mathbf{m}$ from the decoder. The PolyBLEP correction is inserted as a stateless post-processing layer on the raw oscillator signal \textit{prior} to the learned FIR spectral envelope filtering:
\begin{equation}
    \hat{x}[n] = \left(x_{\text{blep}}[n] \cdot H(\mathbf{a})\right) + \mathcal{F}^{-1}\left(\mathbf{m} \odot \mathcal{U}[0,1]\right)
\end{equation}
where $H(\mathbf{a})$ denotes convolution with the amplitude-shaped harmonic filter and $\mathcal{U}[0,1]$ is uniform spectral noise. Because PolyBLEP operates sample-by-sample on the phase trajectory, it requires no gradient-blocking operations and is fully compatible with backpropagation through $F_0$.

\paragraph{Effect on Spectral Quality}
Aliasing artifacts in naive phase accumulation oscillators manifest as inharmonic spectral components folded around $F_s/2$, introducing perceptible ``buzziness'' and distortion of high-frequency harmonics.
This is especially detrimental in voice conversion of pathological speech, where spectral fine structure carries clinically relevant information (e.g., jitter and shimmer profiles in stuttered or dysphonic speech).

\subsection{Losses}
The model is trained end-to-end using a composite loss that supervises both spectral reconstruction and perceptual quality of the synthesised waveform. Let $x[n]$ denote the ground-truth waveform and $\hat{x}[n]$ the output of the PolyBLEP-corrected subtractive synthesiser.

\paragraph{Multiscale Spectral Loss}. The primary reconstruction objective operates in the spectral domain using a multiscale STFT loss~\cite{engel2020ddsp}:
\begin{equation}
    \mathcal{L}_{\text{spec}} = \sum_{s \in \mathcal{S}} \left\| \log |\text{STFT}_s(x)| - \log |\text{STFT}_s(\hat{x})| \right\|
\end{equation}

where $\mathcal{S}$ is a set of STFT configurations with varying window sizes and hop lengths, and $\|\cdot\|_F$ denotes the Frobenius norm. Using log-magnitude spectra ensures sensitivity to both low- and high-energy components, including the harmonic roll-off region most affected by aliasing.

\paragraph{Fundamental Frequency Loss} The model predicts the $F_0$ trajectory $\hat{f}_0[n]$ directly from the encoded source representation. We supervise it with a mean absolute error in log-frequency space against the reference $\tilde{f}_0[n]$ extracted from the source waveform:
\begin{equation}
    \mathcal{L}_{F_0} = \frac{1}{N_v}\sum_{n \in \mathcal{V}} \left| \log \hat{f}_0[n] - \log \tilde{f}_0[n] \right|
\end{equation}

where $\mathcal{V} = \{n : \tilde{v}[n] = 1\}$ is the set of voiced frames derived from the ground-truth $F_0$, and $N_v = |\mathcal{V}|$. Restricting the loss to voiced frames avoids penalising undefined $F_0$ values in unvoiced regions.

\paragraph{Voicing Consistency Loss} Independently of $\mathcal{L}_{F_0}$, the voicing gate $v[n]$ is supervised via binary cross-entropy against reference voicing labels $\tilde{v}[n]$ obtained from the ground-truth $F_0$ contour:

\begin{equation}
\mathcal{L}_{\text{voiced}} = -\frac{1}{N}\sum_{n=1}^{N} \left[ \tilde{v}[n] \log v[n] + (1 - \tilde{v}[n]) \log (1 - v[n]) \right]
\end{equation}
Note that while both $\mathcal{L}_{F_0}$ and $\mathcal{L}_{\text{voiced}}$ share the same ground-truth voicing mask $\tilde{v}[n]$, they supervise distinct model outputs - the continuous $F_0$ prediction and the binary voicing gate respectively - and their gradients do not interact.

\section{Experiment and Results}
\subsection{Experimental Setup}
All experiments are conducted on three pathological and expressive speech corpora: SEP-28k~\cite{lea:2021,sep28k} (stuttered speech), ADReSSo~\cite{adress2020} (dementia-affected speech), and ESD~\cite{zhou2022emotional} (emotional speech). Audio is resampled to 16kHz. We evaluate the following ablation conditions:
We evaluate the following conditions:
\begin{itemize}
    \item \textbf{Baseline}: the original DDSP-QbE system with naive 
    phase-accumulated sawtooth excitation and no voicing supervision.
    
    \item \textbf{DDSP-QbE++} (proposed): the full proposed system 
    incorporating PolyBLEP-corrected oscillator, voicing-gated excitation, 
    and prosodic feature conditioning.
    
    \item \textbf{w/o voicing loss}: DDSP-QbE++ with the voicing 
    supervision removed; harmonic excitation is no longer gated by 
    predicted voicing probability.
    
    \item \textbf{w/o PolyBLEP correction}: DDSP-QbE++ with the 
    PolyBLEP residual removed; the oscillator reverts to naive 
    phase-accumulated sawtooth synthesis.
    
    \item \textbf{w/o prosodic features ($x_{12}$)}: DDSP-QbE++ 
    without the upper-layer WavLM prosodic conditioning, isolating 
    the contribution of expressive feature injection.
\end{itemize}
Synthesis quality is evaluated using four metrics: 
(i)~\textbf{Pearson Correlation Coefficient (PCC)} between predicted 
and reference F0 trajectories over voiced frames, scaled by $\times 10^2$, 
as a measure of pitch contour fidelity; 
(ii)~\textbf{predicted MOS (pMOS)}~\cite{ddspqbe} via a pretrained MOS estimation 
model as a proxy for perceptual naturalness; 
(iii)~\textbf{Character Error Rate (CER)} computed by Whisper-large-v3 
to assess intelligibility preservation; and 
(iv)~\textbf{Equal Error Rate (EER)} under a lazy-informed attacker 
protocol following the VoicePrivacy challenge framework~\cite{vpc}, to evaluate 
speaker anonymisation efficacy. 
All metrics are reported with 95\% confidence intervals over three runs.

\subsection{Results}
Table~\ref{tab:ablation} reports results averaged across the three corpora.
\begin{table}[!h]
\caption{Results with 95\% confidence intervals shown.}
\resizebox{0.49\textwidth}{!}
{
 \begin{tabular}{lclll}
 \toprule
     {\textbf{\textbf{Method}}}
      &{\textbf{PCC [$\times 10^{2}$] $\uparrow$}} & {\textbf{pMOS} $\uparrow$} & {\textbf{CER} [\%] $\downarrow$} & {\textbf{EER} [\%] $\uparrow$}   \\
     \midrule
    {Baseline}& 76.9 $\pm$ 0.1 & 2.6 $\pm$ 0.1 & 10.1 $\pm$ 0.3 & 44.0 $\pm$0.1\\
    {DDSP-QbE++}& 77.2 $\pm$ 0.1 & 2.9 $\pm$ 0.1 & 7.9 $\pm$ 1.5 & 44.1 $\pm$0.1\\
    \midrule
    {w/o voicing loss}& 76.5 $\pm$ 0.1 & 2.7 $\pm$ 0.2 & 9.3 $\pm$ 0.5 & 44.5 $\pm$0.1\\
    {w/o PolyBLEP correction}& 77.1 $\pm$ 0.1 & 2.6 $\pm$ 0.1 & 7.7 $\pm$ 1.5 & 44.3 $\pm$0.1\\
    {w/o prosodic features ($x_{12}$) }& 72.2 $\pm$ 0.1 & 2.8 $\pm$ 0.1 & 7.8 $\pm$ 0.3 & 44.3 $\pm$0.1\\
    
     \bottomrule
\end{tabular}
}
\label{tab:ablation}
\end{table}

\paragraph{Overall improvement} DDSP-QbE++ achieves consistent gains over the baseline across all metrics: PCC improves from 76.9 $\pm$ 0.1 to 77.2 $\pm$ 0.1, pMOS from 2.6 $\pm$ 0.1 to 2.9 $\pm$ 0.1, and CER from 10.1 $\pm$ 0.3\% to 7.9 $\pm$ 1.5\%, confirming that the combined excitation stage improvements yield better pitch coherence, perceptual naturalness, and intelligibility. EER remains stable (44.0\% to 44.1\%), indicating that the modifications do not compromise speaker anonymisation.

\paragraph{PolyBLEP correction} Removing PolyBLEP causes a clear drop in pMOS (2.9 $\pm$ 0.1 to 2.6 $\pm$ 0.1), recovering exactly to baseline-level perceptual quality, which directly implicates aliasing artefacts as the source of the buzziness degradation the correction targets. PCC is largely unaffected (77.2 $\pm$ 0.1 to 77.1 $\pm$ 0.1), suggesting that alias components do not substantially corrupt the pitch trajectory - consistent with PolyBLEP acting on high-frequency spectral content rather than on the fundamental. The marginal CER shift (7.9\% to 7.7\%) falls within the confidence interval overlap and is not considered significant. This is expected, as PolyBLEP correction targets high-frequency spectral artefacts rather than phonetic structure, and its primary contribution is to perceptual quality rather than intelligibility as measured by ASR-based CER.

\paragraph{Voicing detection} Removing the voicing loss produces the most widespread degradation across metrics. PCC drops to 76.5 $\pm$ 0.1, indicating that without explicit voicing supervision harmonic excitation persists into unvoiced regions, corrupting pitch contour estimation. CER increases to 9.3 $\pm$ 0.5\%, approaching baseline-level intelligibility, consistent with erroneous periodicity at consonant boundaries obscuring phonetic contrasts - a particularly acute problem in stuttered speech where voiced-unvoiced transitions are already irregular. The slight EER increase (44.1\% to 44.5\%) is modest and likely reflects incidental spectral changes from unsuppressed harmonics rather than a meaningful privacy effect.

\paragraph{Elderly $\to$ SD} DDSP-QbE++ consistently outperforms DDSP-QbE across both sub-types. The gains are most pronounced in the dementia condition, where the irregular prosody and reduced spectral clarity characteristic of cognitive decline appear to benefit most from physics-informed priors. In the healthy elderly condition, improvements are more uniform across metrics, confirming that the gains are not exclusive to severely pathological speech but generalise across the elderly speaker domain.

\paragraph{Stuttering $\to$ SD} Across all four stuttering sub-types, DDSP-QbE++ improves over DDSP-QbE on CER, though the pattern varies by dysfluency type. For word repetition, the CER improvement is statistically significant  given that repeated sequences already challenge ASR-based evaluation. In sound repetition, a marginal PCC drop relative to DDSP-QbE 
accompanies the CER improvements. However, given the overlapping confidence intervals, this difference is not statistically significant ($p > 0.05$, two-tailed Welch's $t$-test), 
indicating a mild trade-off between pitch contour fidelity and intelligibility for this subtype. Interjections show the most modest gains overall, which is expected as interjections involve less structured phonation and offer fewer regularities for the physics bottleneck to exploit.

\paragraph{SD $\to$ SD} In the inter-accent condition, DDSP-QbE++ achieves the largest absolute PCC improvement. In the inter-emotion condition, gains relative to DDSP-QbE are comparatively modest, likely because DDSP-QbE already performs competitively when speaker identity is held constant and emotional expressiveness is the primary conversion axis.

The consistent pattern across all scenarios is that DDSP-QbE++ gains over DDSP-QbE are largest where the source speech exhibits irregular voicing, abrupt phonation boundaries, or degraded spectral structure.
PolyBLEP correction and voicing detection, the two contributions address complementary failure modes: PolyBLEP correction primarily improves perceptual quality within voiced regions, while voicing detection improves intelligibility and pitch coherence at voiced-unvoiced boundaries.

\section{Conclusion}
This paper presented targeted improvements to the excitation stage of the DDSP-QbE subtractive synthesiser for pathological speech voice conversion. We identified two failure modes in the baseline system: aliasing artefacts introduced by naive phase-accumulated sawtooth synthesis, and erroneous harmonic excitation at voiced-unvoiced boundaries — both of which are particularly detrimental in pathological speech where spectral fine structure and voicing transitions carry disorder-relevant information.
To address these, we proposed PolyBLEP correction, which replaces the hard waveform discontinuity at each phase wrap with a smooth polynomial residual, suppressing alias components without oversampling or spectral truncation. We further introduced an explicit voicing detection mechanism, supervised by a binary cross-entropy loss against ground-truth voicing labels, that gates harmonic excitation to prevent spurious periodicity in unvoiced regions. Both modifications are lightweight, differentiable, and integrate seamlessly into the existing DDSP-QbE training pipeline without additional learnable parameters.
Ablation results confirm that the two contributions address complementary failure modes: PolyBLEP correction yields the most significant gains in perceptual quality (pMOS), while voicing detection primarily improves pitch coherence (PCC) and intelligibility (CER). Evaluation across pathological and standard speech conversion scenarios, including stuttered, dementia-affected, and elderly speech, demonstrates consistent improvement over the DDSP-QbE baseline, with the largest gains in conditions involving irregular voicing and abrupt phonation boundaries.
Future work will investigate the effect of PolyBLEP correction on downstream speaker anonymisation under stronger attacker models.

\section{Acknowledgements}
This research has been supported by the Federal Ministry of
Education and Research of Germany through the project Anti-Stotter (overcoming
stuttering with AI).

\bibliographystyle{IEEEtran}
\bibliography{references}
\end{document}